\documentclass[sigconf]{acmart}

\AtBeginDocument{%
  }

\setcopyright{acmlicensed}
\copyrightyear{2024}
\acmYear{2024}
\acmDOI{XXXXXXX.XXXXXXX}

\acmConference[Conference acronym RecSys]{Make sure to enter the correct
  conference title from your rights confirmation emai}{June 03--05,
  2024}{Woodstock, NY}

\acmISBN{978-1-4503-XXXX-X/18/06}

\copyrightyear{2024}
\acmYear{2024}
\setcopyright{rightsretained}
\acmConference[RecSys '24]{18th ACM Conference on Recommender Systems}{October 14--18, 2024}{Bari, Italy}
\acmBooktitle{18th ACM Conference on Recommender Systems (RecSys '24), October 14--18, 2024, Bari, Italy}\acmDOI{10.1145/3640457.3688034}
\acmISBN{979-8-4007-0505-2/24/10}

\begin{document}

\title{Joint Modeling of Search and Recommendations Via an Unified Contextual Recommender (UniCoRn)}

\author{Moumita Bhattacharya}
\authornote{Both authors contributed equally to this research.}
\email{mbhattacharya@netflix.com}
\orcid{1234-5678-9012}
\affiliation{%
  \institution{Netflix Research}
  \city{Los Gatos}
  \state{CA}
  \country{USA}
}

\author{Vito Ostuni}
\authornotemark[1]
\email{vostuni@netflix.com}
\affiliation{%
  \institution{Netflix Research}
  \city{Los Gatos}
  \state{CA}
  \country{USA}
}

\author{Sudarshan Lamkhede}
\email{slamkhede@netflix.com}
\affiliation{%
  \institution{Netflix Research}
  \city{Los Gatos}
  \state{CA}
  \country{USA}
}

\renewcommand{\shortauthors}{Bhattacharya and Ostuni et al.}

\begin{abstract}
Search and recommendation systems are essential in many services, and they are often developed separately, leading to complex maintenance and technical debt. In this paper, we present a unified deep learning model that efficiently handles key aspects of both tasks. 
 

\end{abstract}



\maketitle

\section{Introduction}
In real-world applications, teams often develop separate models to solve search and recommendation tasks. Throughout various services, it is common to have query-driven item searches, item-to-item similarity-based recommendations as well as other kinds of more traditional recommendations. It is often the case that teams develop bespoke models for each use case, which can rapidly result in systems management overhead and hidden technical debt in maintaining a large number of specialized models. As observed by \cite{sculley2015hidden, menezes2023lessons}, this complexity can lead to increased long-term costs, and reduced reliability and effectiveness of ML systems. Moreover, we argue that these different applications can benefit from each other \cite{zamani2018joint}. In this talk, we will describe a series of practical solutions and modeling approaches that we built to leverage one single deep learning model to serve both search and certain recommendations tasks. Additionally, we share approaches that we took to personalize search results at scale, while also improving the recommendation use cases served by this unified model.

\begin{figure*}[t!]
  \centering
  \includegraphics [width=0.83\linewidth]{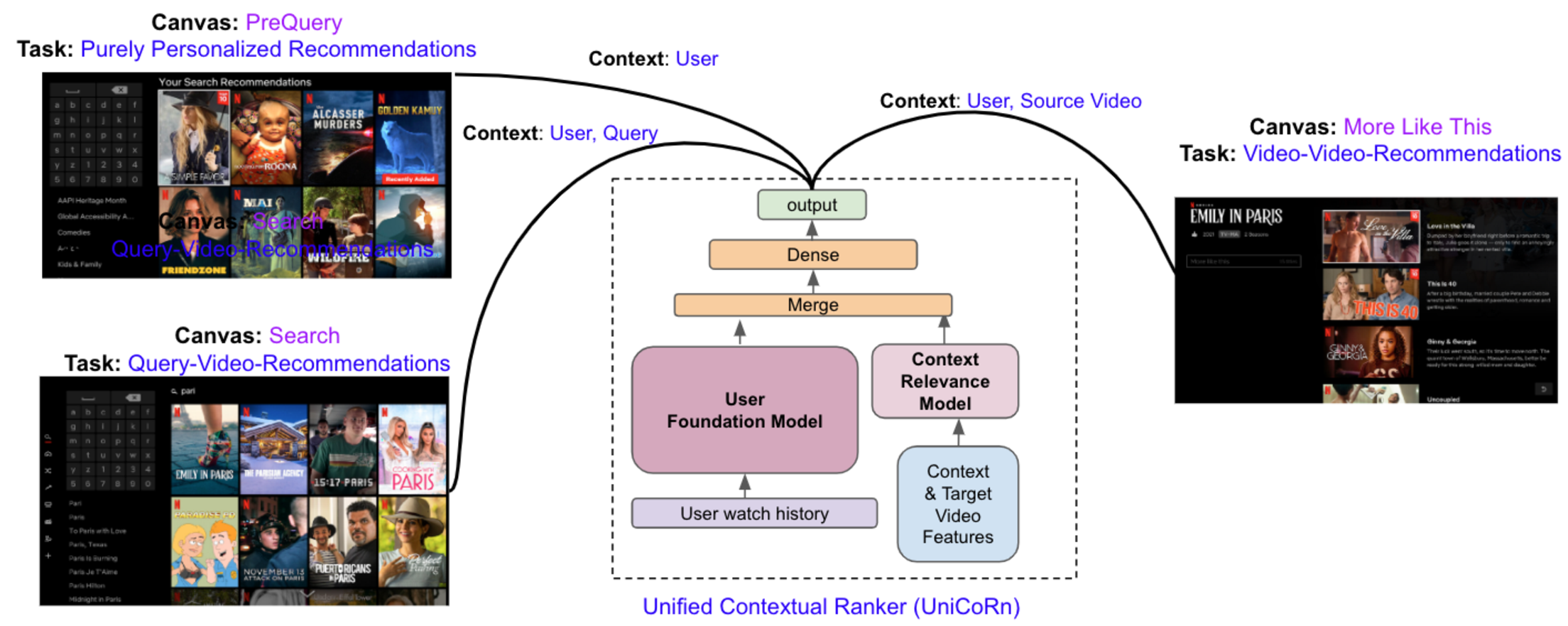}
  \caption{Unified Contextual Ranker (UniCoRn) powering multiple different search and recommendation tasks}
  \label{fig:overview}
\end{figure*}

\section{Proposed Approach}

\textbf{\underline{Model Unification}:} Prior to developing the approach presented here, we had several models powering different applications on the Netflix product. With our approach we were able to unify some of these models. For example, we trained a model that exclusively powered Netflix Search canvas (\textit{Query-Video-Recommendations}), where when a member types in a query, we show videos that are relevant to the query \cite{Netflix-Search}. Similarly, we trained a separate model that generated recommendations for a given video when a user clicks on it to show other videos that are similar to it (\textit{Video-Video-Recommendations}). We trained yet another model to power recommendations on PreQuery or anticipatory search canvas (\textit{Profile-Video-Recommendations} \cite{Bhattacharya-recsys-2022}). As we consider search and recommendation as "two sides of the same coin", we embarked into a journey of unification of all these different models, such that not only we consolidate the tech stack, but also be able to use just one trained model to serve all these different types of applications across different parts of the Netflix product, as shown in Fig. \ref{fig:overview}. 

One of the key differences between a traditional search ranking model and a traditional recommender system is the input context. Search is centered around an explicit textual \textit{query} context, while recommendations are driven by the \textit{user context} and/or other context such as \textit{source item}. More specifically, when a user visits the Search PreQuery page they are shown a list of recommendations that are personalized for them. As soon as the user types something, they are shown search results relevant to their query. We approached these two separate context driven tasks as one task with a \textbf{shared broader context definition}. 

We develop a model that has in its \textbf{context} the following information: \textit{user id},  \textit{query}, \textit{country}, \textit{source entity id} and \textit{task}. Note, \textit{entity id} here refers to the \textit{id} of a video or a game. The \textbf{output} of this model is a probability score for positive engagement with an entity, which we will refer to as \textit{target entity id} from here on. This model is trained on a dataset that is gathered from engagements pertaining to all the different tasks, which was possible because of the broadened context.

For some tasks, only certain contexts are available. For example, for the search task, we have  \textit{query}, \textit{user id}, \textit{country} and \textit{task} contexts available but we don’t have any \textit{source entity id} information. Similarly, for Video-Video Recommendations tasks we don’t have \textit{query} in the context. We developed different heuristics to impute missing contexts for each tasks. Specifically, for search tasks we impute null value to the missing context, whereas for certain recommendations tasks such video-video or entity-video recommendations, we impute the missing  \textit{query} context by leveraging tokens of the display names of the entity. This imputation approach reduces missing context, which helps with improved cross-application learning. 

We use several features, which can be broadly classified into two types: (1) context specific features (such as query length, source entity id embedding) and (2) context and \textit{target entity id} features (such as number of clicks of the \textit{target entity id} for a given query). These features are either categorical or real valued numeric features. We trained a deep learning model, where all these features are fed in the input layer and each categorical feature is leveraged to learn corresponding embedding layers. The model architecture includes residual connections and feature crossing. We use a binary cross entropy loss, with an Adam optimizer.

Once trained, this model then can generate different ranked results for different types of context. For example, the same model can show videos for a query, while on PreQuery recommend personalized videos for the user. In fact, after a series of experiments and improvements we have one model that currently powers \textit{Netflix Search}, \textit{Personalized Pre-query canvas}, \textit{More Like this Canvas}, among others. We will refer to this proposed unified contextual model as \textit{UniCoRn} from here on. With this unification we were able to achieve either a lift or parity in performance for different tasks.

Our work shows how training one model on a larger dataset and having individual tasks share data among them performs better than the individual models trained on their own, as each task benefits from auxiliary ones. A few things in our setup that enables such information passing among the different tasks, thus making it possible for us to serve both search and recommendations tasks via one model are: (1) adding the task type as context and having features that are specific to different tasks helps the model to learn trade-offs between the different tasks. (2) Imputing in missing contexts wherever possible helps better feature coverage and enable the model to learn across different tasks better. (3) Feature crossing also helps with cross task learning.

\textbf{\underline{Personalization of the Unified Model}:} A single model powering both Search and Recommendation tasks, gives us the advantage of easily leveraging the personalization capabilities typical of the Recommendation task in the Search task. However, blindly applying personalization to Search may incur undesirable consequences where personalization takes over query relevance. Therefore, we need to control for context relevance and personalization accordingly. A fully personalized search experience also poses serving challenges as we need to meet strict latency requirements. The low latency constraint is due to the nature of our instant results experience where we return results for each keystroke \cite{NetflixCaseStudy}. 

We took an incremental approach to personalize the unified model. We started with a semi-personalized model based on user clustering, where the user cluster assignment is a context level feature. Although this experience was not fully personalized, it had the advantage of  still enabling results caching. We then moved to a more powerful fully personalized model, first relying on separate recommendation model's outputs used as features. More details about these approaches are in \cite{Vito_netflix_search_personalization}. Subsequently, we developed an end-to-end architecture, which incorporated a pre-trained user and item representations model that was fine-tuned with UniCoRn. These personalization approaches with appropriate model architecture led to significant improvements in the offline metrics for both search and recommendations tasks. Specifically, the lift from a vanilla non-personalized UniCoRn to a fully personalized one was ~7\% and ~10\% for search and a recommendations, respectively. 

\section{Conclusion} This work demonstrates that a single unified model, aware of diverse contexts, can perform and improve both search and recommendation tasks. Additionally, incorporating personalization within this model benefits both applications, while optimally trading-off relevance and personalization.

\section{Authors Bio}
Moumita Bhattacharya and Vito Ostuni are research scientist at Netflix where they work on Search and Recommendation algorithms. Sudarshan Lamkhede is a Engineering Manager at Netflix leading the Foundation Model and Search Algorithms team. 

\begin{acks}
We are thankful to our collaborators Roger Menezes, Gary Yeh, Manjesh Nilange, Jinning Zhong, Guru Tahasildar, Christoph Kofler and Raveesh Bhalla as well as internal reviewer Justin Basilico.
\end{acks}

\bibliographystyle{ACM-Reference-Format}
\bibliography{references}

\end{document}